\documentclass[]{pasj01}

\begin{document} 
\Received{}
\Accepted{}

\title{~Localized Recombining Plasma in G166.0+4.3:
A Supernova Remnant with an Unusual Morphology}

\author{Hideaki \textsc{Matsumura}\altaffilmark{1}%
}
\altaffiltext{1}{Department of Physics, Faculty of Science, Kyoto University, Kitashirakawa Oiwake-cho, Sakyo-ku, Kyoto-shi, Kyoto 606-8502, Japan}

\author{Hiroyuki \textsc{Uchida} \altaffilmark{1}}

\author{Takaaki \textsc{Tanaka} \altaffilmark{1}}

\author{Takeshi~Go \textsc{Tsuru} \altaffilmark{1}}

\author{Masayoshi \textsc{Nobukawa} \altaffilmark{2}}
\altaffiltext{2}{Science, Curriculum and Instruction, Faculty of Education, Nara University of Education, Takabatake-cho, Nara-shi, Nara 630-8528, Japan}

\author{Kumiko~Kawabata \textsc{Nobukawa} \altaffilmark{3}}
\altaffiltext{3}{Department of Physics, Nara Women's University, Kitauoyanishimachi, Nara 630-8506, Japan}


\author{Makoto \textsc{Itou} \altaffilmark{1}}



\KeyWords{X-rays: individual, G166.0+4.3, Recombining plasma} 

\maketitle

\begin{abstract}
We observed the Galactic mixed-morphology supernova remnant G166.0+4.3 with Suzaku.
The X-ray spectrum in the western part of the remnant is well represented by a one-component ionizing plasma model.
The spectrum in the northeastern region can be explained by two components.
One is the Fe-rich component with the electron temperature $kT_e = 0.87_{-0.03}^{+0.02}~\rm keV$. 
The other is the recombining plasma component of lighter elements with $kT_e = 0.46\pm0.03~\rm keV$, the initial temperature $kT_{init} = 3~\rm keV~(fixed)$ and the ionization parameter $n_et = (6.1_{-0.4}^{+0.5}) \times 10^{11}~\rm cm^{-3}~s$.
As the formation process of the recombining plasma, two scenarios, the rarefaction and thermal conduction, are considered.
The former would not be favored since we found the recombining plasma only in the northeastern region whereas the latter would explain the origin of the RP.
In the latter scenario, an RP is anticipated in a part of the remnant where blast waves are in contact with cool dense gas.
The emission measure suggests higher ambient gas density in the northeastern region.
The morphology of the radio shell and a GeV gamma-ray emission also suggest a molecular cloud in the region.

\end{abstract}

\section{Introduction}
Mixed-morphology supernova remnants (MM SNRs) have center-filled thermal X-ray emissions in a synchrotron radio shell \citep{Rho1998}.
Although the formation mechanism of the non-standard morphology has not been fully understood, there are two proposed scenarios. 
One is the evaporating cloudlet scenario \citep{White1991}, and the other is the thermal conduction scenario \citep{Cox1999,Shelton1999}.
In either scenario, interactions between the blast wave and dense gas play an important role in the formation of the X-ray morphology. 
Most MM SNRs are actually associated with molecular clouds as indicated by CO line emissions or OH (1720 MHz) masers (e.g., \cite{Lazendic2006}).
They also exhibit GeV/TeV gamma-ray emissions which suggest shell-cloud interactions since those SNRs can emit luminous gamma rays through $\pi^{0}$ decays (e.g., \cite{Ackermann2013}; \cite{Albert2007}).

In the ASCA observation of the MM SNR IC~443, Kawasaki et al. (2002) measured ionization temperatures ($T_{z}$) based on the H-like to He-like K$\alpha$ intensity ratios of Si and S, and compared them with electron temperatures ($T_{e}$).
They found that $T_{z}$ is significantly higher than $T_{e}$, and suggested that the plasma is overionized.
In the observations of IC~443 \citep{Yamaguchi2009} and W49B \citep{Ozawa2009} with Suzaku, they discovered radiative recombining continua (RRC), which provide clear evidence that the plasmas are in an extremely recombining state.
Subsequent Suzaku observations revealed recombining plasmas (RPs) in other MM SNRs in the Galaxy (e.g., G359.1$-$0.5: \cite{Ohnishi2011}; W28: \cite{Sawada2012}; W44: \cite{Uchida2012}) as well as in the Large Magellanic Cloud (N49: \cite{Uchida2015}).

Two scenarios are mainly considered for the formation process of the RPs.
One is the rarefaction scenario \citep{Itoh1989}.
If a supernova explodes in a dense circumstellar matter (CSM) around the massive progenitor, ejecta and the CSM are shock-heated and quickly ionized because of the high density.
When the shock breaks the CSM out to a lower density interstellar medium (ISM), $T_e$ is decreased by adiabatic cooling.
\citet{Kawasaki2002} proposed the other scenario, the thermal conduction scenario.
If an SNR shock is interacting with a molecular cloud, thermal conduction occurs between the SNR plasma and the cloud, and, therefore, $T_e$ drops.
Since the recombination timescales of ions are generally longer than the conduction timescales, $T_{z}$ cannot follow the decrease of $T_e$, and RPs can be realized.

G166.0+4.3 is a Galactic SNR whose radio shell has a large bipolar structure in the west with a smaller semicircle shell in the east \citep{Sharpless1959,Landecker1982,Pineault1987}.
\citet{Landecker1982} estimated the distance to G166.0+4.3 as $5.0\pm0.5$~kpc by using the $\Sigma$-$D$-$z$ (surface brightness-diameter-distance above the Galactic plane) relationship.
\citet{Pineault1987} suggested that the gas density in the east is higher than that in the west because of the unusual morphology of the radio shell.
G166.0+4.3 was observed in the X-ray band with ROSAT \citep{Burrows1994}, ASCA \citep{Guo1997} and XMM-Newton \citep{Bocchino2009}.
\citet{Burrows1994} found the X-ray morphology is distinct from the radio morphology and reproduced the spectrum with collisional ionization equilibrium (CIE: $T_{z} = T_{e}$) models.
They estimated the age of the remnant to be $\sim 2.4\times10^{4}$~yr by applying the evaporating cloudlet model \citep{White1991} when they assume the estimated distance of 5~kpc.
\citet{Bocchino2009} classified G166.0+4.3 as an MM SNR and reported that the spectra can be reproduced by an ionizing plasma (IP: $T_{z} < T_{e}$) model.

Gas environment of G166.0+4.3 is clearly different between the east and the west, according to the morphology of the radio shell.
It is one of the best sources to study plasma evolution depending on the environment.
In this paper, we report on observational results of G166.0+4.3 with Suzaku, which has high sensitivity and high energy resolution in the energy band of 0.3--12~keV.
We assume $5~{\rm kpc}$ as the distance to the remnant in this paper.

\section{Observations and data reduction}
We performed Suzaku observations of G166.0+4.3.
Table~\ref{tab:obs_log} summarizes the observation log.
In this analysis, we used data from the X-ray Imaging Spectrometer (XIS: \cite{Koyama2007}) installed on the focal planes of the X-ray telescopes (XRT: \cite{Serlemitsos2007}).
XIS2 and a part of XIS0 are not used in what follows.
The former has not been functioning since 2006 November.
The latter was turned off in 2009 June (Suzaku XIS documents\footnote{http://www.astro.isas.ac.jp/suzaku/doc/suzakumemo/suzakumemo-2007-08.pdf}$^{,}$\footnote{http://www.astro.isas.ac.jp/suzaku/doc/suzakumemo/suzakumemo-2010-01.pdf}).

We reduced the data using the HEADAS software version 6.19.
We used the calibration database released in 2015 October for data processing.
The total exposure time of the five observations is $\sim 226$~ks after the standard data screening.
In the processing of the XIS data, we removed cumulative flickering pixels by referring to the noisy pixel maps\footnote{https://heasarc.gsfc.nasa.gov/docs/suzaku/analysis/xisnxbnew.html} provided by the XIS team.
We also discarded pixels adjacent to the flickering pixels (pixel quality $\rm B14=1$). 
Non X-ray backgrounds (NXBs) were estimated by {\tt xisnxbgen} \citep{Tawa2008}.
The redistribution matrix files and the ancillary response files were produced by {\tt xisrmfgen} and {\tt xissimarfgen} \citep{Ishisaki2007}, respectively.

\begin{table*}
  \tbl{Observation log.}{
  \begin{tabular}{lllll}
      \hline
      Target & Obs. ID & Obs.~date & (R.A., Dec.) & Effective Exposure\\      
      \hline
	G166.0+4.3\_NE & 509022010 & 2014-Sep-19 & (\timeform{5h27m07.7s},~\timeform{42D58'52.2''}) & 61~ks\\
	G166.0+4.3\_NE & 509022020 & 2014-Sep-22 & (\timeform{5h27m07.7s},~\timeform{42D58'52.2''}) & 61~ks\\
	G166.0+4.3\_NW & 509023010 & 2014-Sep-20 & (\timeform{5h25m46.8s},~\timeform{42D54'01.7''}) & 42~ks\\
	G166.0+4.3\_SE & 509024010 & 2014-Sep-21 & (\timeform{5h26m41.2s},~\timeform{42D39'06.2''}) & 34~ks\\
	G166.0+4.3\_SE & 509024020 & 2015-Mar-13 & (\timeform{5h26m41.2s},~\timeform{42D39'06.2''}) & 27~ks\\
	IRAS 05262+4432 (Background) & 703019010 & 2008-Sep-14 & (\timeform{5h29m56.0s},~\timeform{44D34'39.2''}) & 82~ks\\
      \hline
    \end{tabular}}
    \label{tab:obs_log}
\begin{tabnote}

\end{tabnote}
\end{table*}

\section{Analysis}

\subsection{Image}
\begin{figure}
\begin{center}
 \includegraphics[width=7cm]{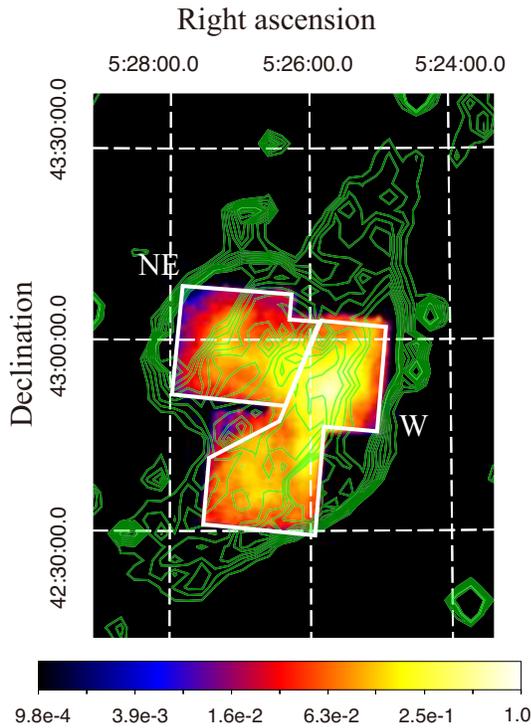} 
\end{center}
\caption{XIS image of G166.0+4.3 in the 0.5--2.0 keV band after the NXB subtraction and the correction of the vignetting effect. 
The coordinates refer to epoch J2000.0.
The X-ray count rate is normalized so that the peak becomes unity.
We overlaid contours of the 325 MHz radio image from the WENSS.
}
\label{fig:img}
\end{figure}

Figure~\ref{fig:img} shows a mosaic image of G166.0+4.3 in the energy band of 0.5--2.0~keV.
We subtracted the NXB from the image and corrected it for the vignetting effect of the XRT.
Contours are a 325~MHz radio image from the Westerbord Northern Sky Survey (WENSS).

\subsection{Background subtraction}

For background estimation, we used Suzaku observational data of IRAS~05262+4432, which is located at ($l, b$) = (\timeform{165.1D}, \timeform{5.7D}).
The observation log is shown in Table~\ref{tab:obs_log}.
We extracted a spectrum from a source-free region.
In order to model the spectrum, we adopted a model by \citet{Masui2009}.
They observed the direction ($l, b$) = (\timeform{230D}, \timeform{0D}) with Suzaku in order to study the soft X-ray emission from the Galactic disk.
Their model consists of four components: the cosmic X-ray background (CXB), the local hot bubble (LHB), and two thermal components for the Galactic halo ($\rm GH_{cold}$ and $\rm GH_{hot}$).

\begin{figure}
\begin{center}
 \includegraphics[width=8cm]{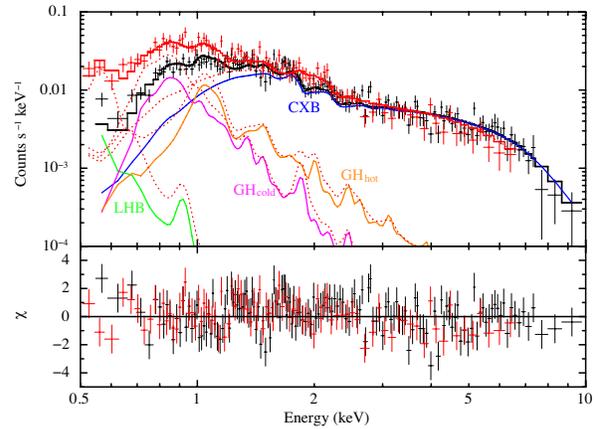} 
\end{center}
\caption{XIS0+3 (black) and XIS1 (red) spectra taken from the source-free region of the IRAS~05262+4432 data with the best-fit model. 
The blue, green, magenta and orange lines represent the CXB, LHB, $\rm GH_{cold}$ and $\rm GH_{hot}$ models, respectively. 
The bottom panel shows the data residuals from the best-fit model.
}
\label{fig:bkg_spe}
\end{figure}

We fit the background spectrum with the model after the NXB subtraction using XSPEC version 12.9.0.
We fixed the electron temperatures of the LHB, $\rm GH_{cold}$ and $\rm GH_{hot}$ at the values by \citet{Masui2009}.
The absorption column density ($N_{\rm H}$) of the CXB was fixed at 3.8$\times$10$^{21}$ cm$^{-2}$, the Galactic value in the line of sight toward IRAS~05262+4432 \citep{Dickey1990,Kalberla2005}.
The normalizations of all the components were allowed to vary.
Figure~\ref{fig:bkg_spe} and Table~\ref{tab:bkg_model} show the fitting result and the best-fit parameters, respectively.

\begin{table}
  \tbl{Best-fit model parameters of the background spectra}{
  \begin{tabular}{llll}
      \hline
      Component & Parameter (unit) & Value \\      
      \hline
      CXB & $N_{\rm H}$ (10$^{21}$cm$^{-2}$) & 3.8 (fixed) \\
      & Photon index & 1.4 (fixed) \\
      & Normalization$^{\ast}$ & 10.7 $\pm$ 0.4 \\
      \hline
      LHB & $kT_e$ (keV) & 0.105 (fixed) \\
      & Normalization$^{\dagger}$ & 13.4 $\pm$ 3.2\\
      \hline
      $\rm GH_{cold}$ & $kT_e$ (keV) & 0.658 (fixed) \\
      & Normalization$^{\dagger}$ & 2.1 $\pm$ 0.2\\
      $\rm GH_{hot}$ & $kT_e$ (keV) & 1.50 (fixed) \\
      & Normalization$^{\dagger}$ & 3.1 $\pm$ 0.5\\
      \hline
      $\chi^{2}$ (d.o.f.) & & 301.9 (234)\\
      \hline
    \end{tabular}}
    \label{tab:bkg_model}
\begin{tabnote}
${\ast}$The unit is photons s$^{-1}$cm$^{-2}$keV$^{-1}$sr$^{-1}$ @ 1~keV.\\
$^{\dagger}$The emission measure integrated over the line of sight, i.e., (1 / 4$\pi$) $\int n_e n_{\rm H} dl$ in units of 10$^{14}$cm$^{-5}$sr$^{-1}$.
\end{tabnote}
\end{table}

\subsection{SNR spectra}
\begin{figure}
\begin{center}
 \includegraphics[width=8cm]{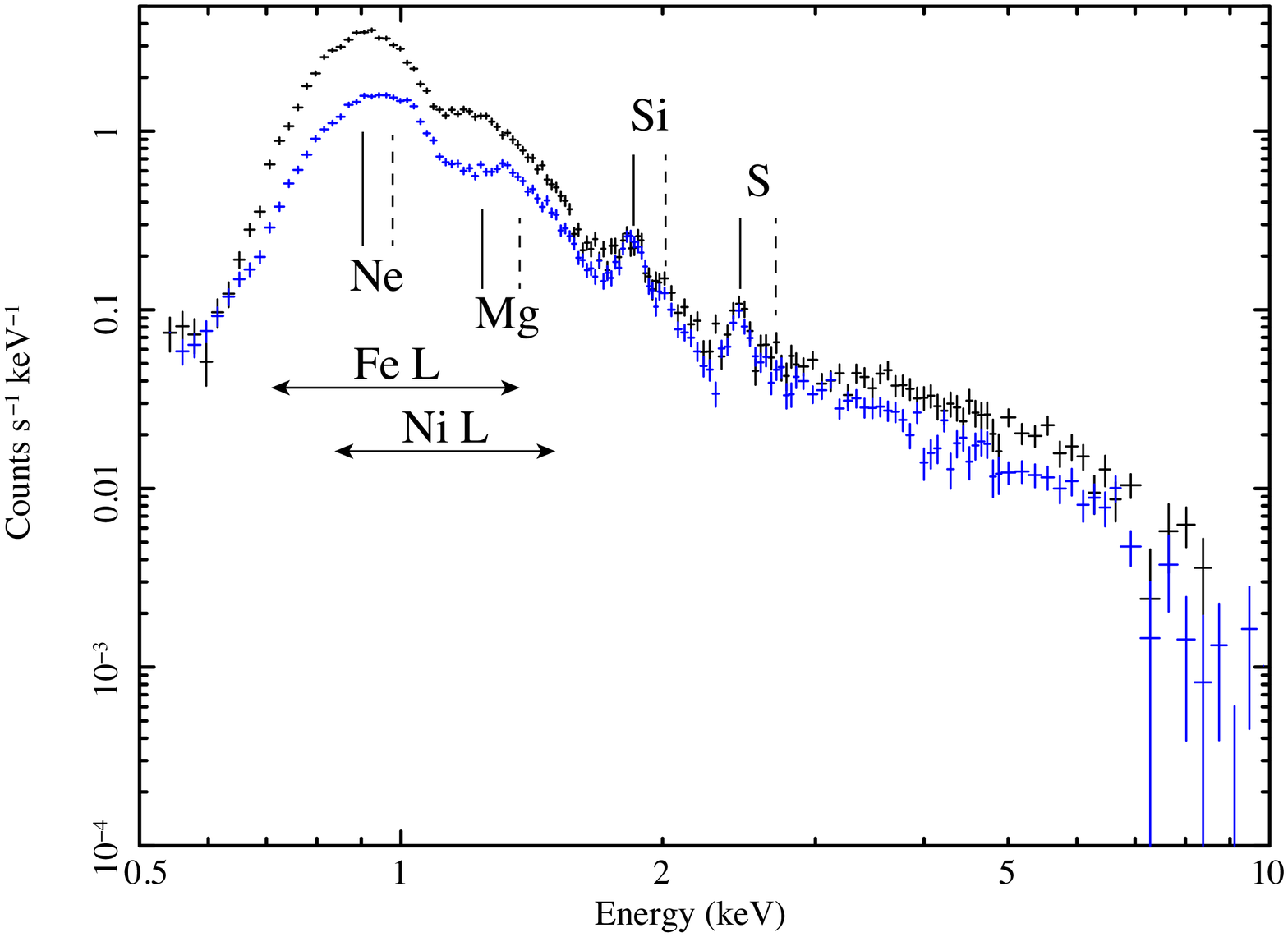} 
\end{center}
\caption{XIS0+3 spectra of W (black) and NE (blue).
The vertical solid and dashed black lines show the center energies of the K lines from He-like and H-like ions, respectively.
The arrows indicate the energy bands with Fe-L and Ni-L line complexes. 
}
\label{fig:raw_spe}
\end{figure}

Considering the morphology of the radio shell, we extracted spectra from the northeast and west regions (named NE and W) indicated with the white solid lines in Figure~\ref{fig:img}.
Figure~\ref{fig:raw_spe} shows XIS0+3 spectra of W and NE after the NXB subtraction.
One can clearly see Fe-L and Ni-L line complexes, and Si-K and S-K lines.
The Fe-L and Ni-L line complexes are more prominent in W than those in NE.

We fitted the spectra with the model consisting of the SNR component and the background ($\S$3.2).
The overall normalization of the background component was treated as a free parameter.
The absorption column density of the CXB component was fixed at 3.6$\times$10$^{21}$ cm$^{-2}$ which is the Galactic value in the line of sight toward G166.0+4.3.

\begin{figure*}
\begin{center}
 \includegraphics[width=16cm]{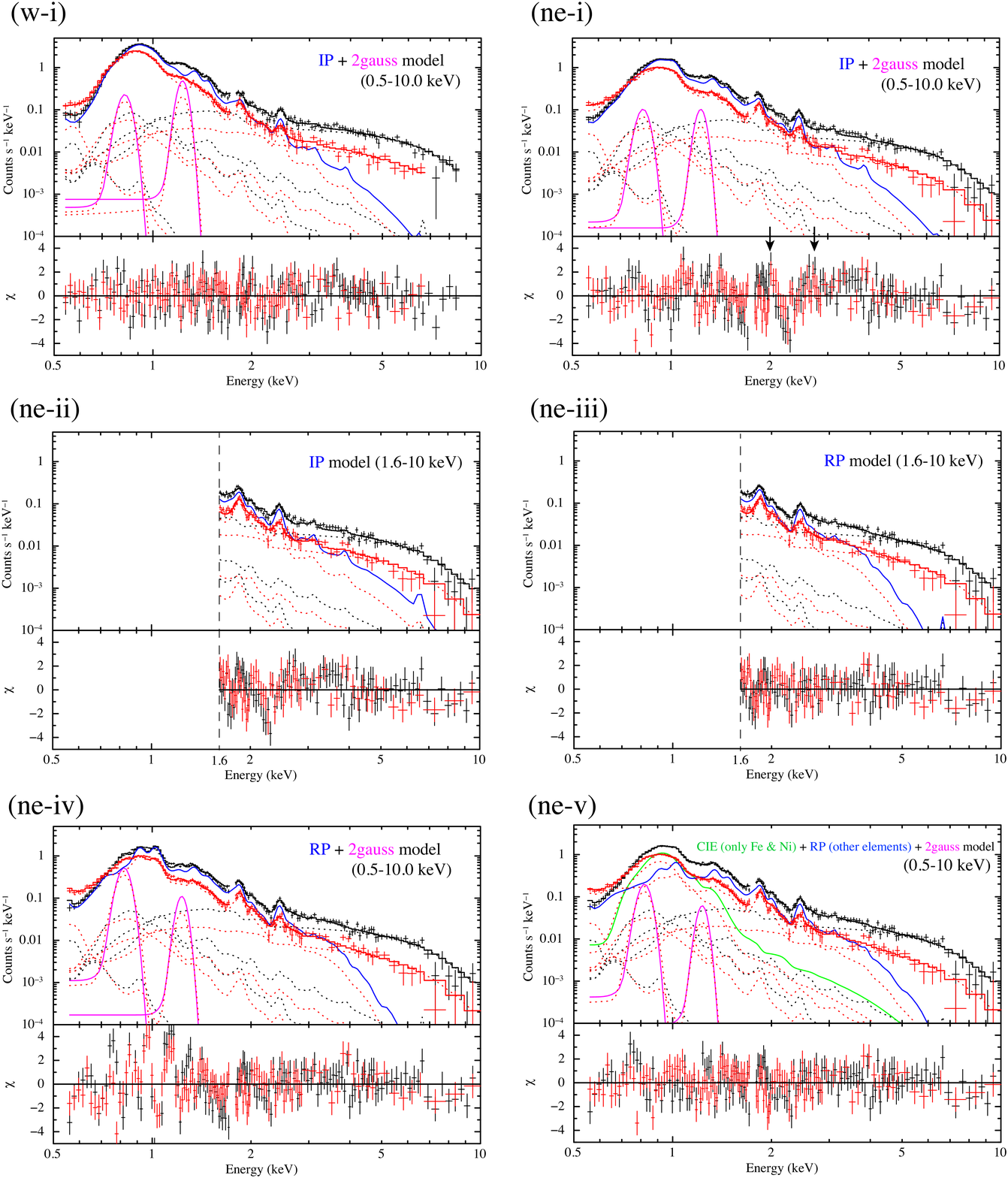} 
\end{center}
\caption{XIS0+3 (black) and XIS1 (red) spectra of W or NE.
Each spectrum is fitted with the plasma model (solid lines) and the background model (dotted lines). 
}
\label{fig:spectra}
\end{figure*}

We applied a one-component IP model to the W spectrum using the VVRNEI model in the XSPEC package, where the initial temperature ($kT_{init}$) is fixed at 0.01~keV whereas the electron temperature, ionization parameter ($n_{e}t$) and normalization are free parameters.
We used the Wisconsin absorption model \citep{Morrison1983} for the absorption whose column density is a free parameter.
The abundances of O, Ne, Mg, Si, S, and Fe are also free parameters.
The Ar and Ca abundances are linked to S, and Ni is linked to Fe.
The abundances of the other elements are fixed to the solar values \citep{Ander1989}.
The energy band around the neutral Si K-edge (1.73--1.78~keV) is ignored because of the known calibration uncertainty.
The fitting left large residuals around 0.82~keV and 1.23~keV due to the lack of Fe-L lines in the model (e.g., \cite{Yamaguchi2011}).
Therefore, we added two narrow lines at 0.82~keV and 1.23~keV.
The results are shown in Figure~\ref{fig:spectra}~(w-i) and Table~\ref{tab:spectra}.
We found that the spectrum is well reproduced by a one-component IP model with the electron temperature of 0.83~keV.

We applied the same IP model as W to the NE spectrum.
Significant residuals are found at $\sim 2.0~\rm keV$ and $\sim 2.6~\rm keV$ which correspond to Si Ly$\alpha$ (2.0 keV) and the edge of the RRC of He-like Si  (2.67~keV) + S Ly$\alpha$ (2.63 keV), respectively (Figure~\ref{fig:spectra}~(ne-i)).
These spectral features, if confirmed, provide evidence that the plasma is in a recombination-dominant state.
We tried a one-component RP model.
However, a satisfactory fit was not obtained.

For a detailed investigation, we restricted the energy band to $>1.6~\rm keV$ where we found the large residuals, and tried one-component IP or RP models.
The absorption column density is fixed to the best-fit value for W, $N_{\rm H} = 0.8 \times 10^{21}~{\rm cm^{-2}}$, whereas $kT_{e}$ and $n_{e}t$ are left free.
The $kT_{init}$ is a free parameter for the RP model, whereas it is fixed at 0.01~keV for the IP model.
The abundances of Si and S are free parameters, and those of Ar and Ca are linked to S.
The residuals were not improved with the IP model (Figure~\ref{fig:spectra}~(ne-ii) and Table~\ref{tab:spectra}).
On the other hand, we successfully fitted the spectrum with the RP model (Figure~\ref{fig:spectra}~(ne-iii) and Table~\ref{tab:spectra}).

\begin{table*}
  \tbl{Fit parameters of source regions.}{
  \begin{tabular}{lllllll}
      \hline
      Model function & Parameter (unit) & (w-i) & (ne-i) & (ne-ii) & (ne-iii) & (ne-v)\\      
      \hline
      Absorption & $N_{\rm H}$ (10$^{21}$cm$^{-2}$) & 0.8 $\pm$ 0.1 & 0.3 $\pm$ 0.1 & 0.8 (fixed) & 0.8 (fixed) & 1.7 $\pm$ 0.1\\
      VVRNEI1 & $kT_e$ (keV) & 0.83 $\pm$ 0.01 & 0.90 $_{-0.02}^{+0.01}$ & 1.36 $_{-0.27}^{+0.51}$ & 0.37 $\pm$ 0.04 & 0.46 $\pm$ 0.03\\
      & $kT_{init}$ (keV) & 0.01 (fixed) & 0.01 (fixed) & 0.01 (fixed) & $\geq$ 2.98 & 3.0 (fixed)\\
      & H (solar) & 1 (fixed) & 1 (fixed) & 1 (fixed) & 1 (fixed) & 1 (fixed)\\
      & He (solar) & 1 (fixed) & 1 (fixed) & 1 (fixed) & 1 (fixed) & 1 (fixed)\\
      & C (solar) & 1 (fixed) & 1 (fixed) & 1 (fixed) & 1 (fixed) & 1 (fixed)\\
      & O (solar) & 0.4 $_{-0.1}^{+0.2}$ & 0.2 $_{-0.1}^{+0.2}$ & 1 (fixed) & 1 (fixed) & 0.02  $_{-0.01}^{+0.03}$\\
      & Ne (solar) & 0.6 $_{-0.1}^{+0.2}$ & 0.6 $_{-0.1}^{+0.2}$ & 1 (fixed) & 1 (fixed) & 0.4 $\pm$ 0.1\\
      & Mg (solar) & 1.2 $\pm$ 0.2 & 0.6 $\pm$ 0.1 & 1 (fixed) & 1 (fixed) & 0.6 $\pm$ 0.1\\
      & Si (solar)  & 0.7 $_{-0.2}^{+0.1}$ & 0.6 $\pm$ 0.1 & 1.4 $_{-0.1}^{+0.3}$ & 2.3 $\pm$ 0.3 & 1.0 $\pm$ 0.1\\
      & S = Ar = Ca (solar) & 1.2 $\pm$ 0.3 & 1.2 $\pm$ 0.2 & 1.8 $\pm$ 0.4 & 3.3 $_{-0.6}^{+0.7}$ & 1.5 $\pm$ 0.2\\
      & Fe = Ni (solar) & 1.2 $\pm$ 0.2 & 0.4 $\pm$ 0.1 & 1 (fixed) & 1 (fixed) & -------------\\
      & $n_{e}t$ (10$^{11}$ cm$^{-3}$s) & 4.0 $_{-0.5}^{+0.7}$ & 3.1 $\pm$ 0.5 & 1.6 $_{-0.7}^{+1.7}$ & 6.2 $\pm$ 0.6 & 6.1 $_{-0.4}^{+0.5}$\\
      & VEM ($\rm 10^{57} cm^{-3}$)$^{\dagger}$ & 2.7 $\pm$ 0.2 & 2.9 $_{-0.3}^{+0.2}$ & 1.2 $_{-0.2}^{+0.3}$ & 6.6 $_{-1.3}^{+1.6}$ & 10.3 $_{-1.0}^{+0.8}$\\
      VVRNEI2 & $kT_e$ (keV) & ------------- & ------------- & ------------- & ------------- & 0.87 $_{-0.03}^{+0.02}$\\
      & Fe = Ni (solar) & ------------- & ------------- & ------------- & ------------- & 0.14  $\pm$ 0.01\\
      & VEM ($\rm 10^{57} cm^{-3}$)$^{\dagger}$ & ------------- & ------------- & ------------- & ------------- & = VVRNEI1\\
      Gaussian & Centroid (keV) & 0.82 (fixed) & 0.82 (fixed) & ------------- & ------------- & 0.82 (fixed)\\
      & Normalization$^{\ddagger}$ & 4.6 $_{-0.7}^{+1.3}$ & 2.1 $_{-0.7}^{+0.8}$ & ------------- & ------------- & 6.7 $_{-1.3}^{+1.5}$\\
      Gaussian & Centroid (keV) & 1.23 (fixed) & 1.23 (fixed) & ------------- & ------------- & 1.23 (fixed)\\
      & Normalization$^{\ddagger}$ & 3.4 $_{-0.2}^{+1.3}$ & 0.8 $\pm$ 0.1 & ------------- & ------------- & 0.7 $\pm$ 0.2\\
      \hline
      $\chi^{2}$ (d.o.f.) & & 339.2 (247) & 443.9 (251) & 232.7 (150) & 172.1 (150) & 347.6 (250)\\
      \hline
    \end{tabular}}
    \label{tab:spectra}
\begin{tabnote}
$^{\dagger}$Volume emission measure VEM = $\int n_e n_{\rm H} dV$ , where $n_e$, $n_{\rm H}$, and $V$ are the electron and hydrogen densities, and the emitting volume, respectively.\\
$^{\ddagger}$The unit is photons s$^{-1}$cm$^{-2}$sr$^{-1}$.
\end{tabnote}
\end{table*}

\begin{figure}
\begin{center}
 \includegraphics[width=6.5cm]{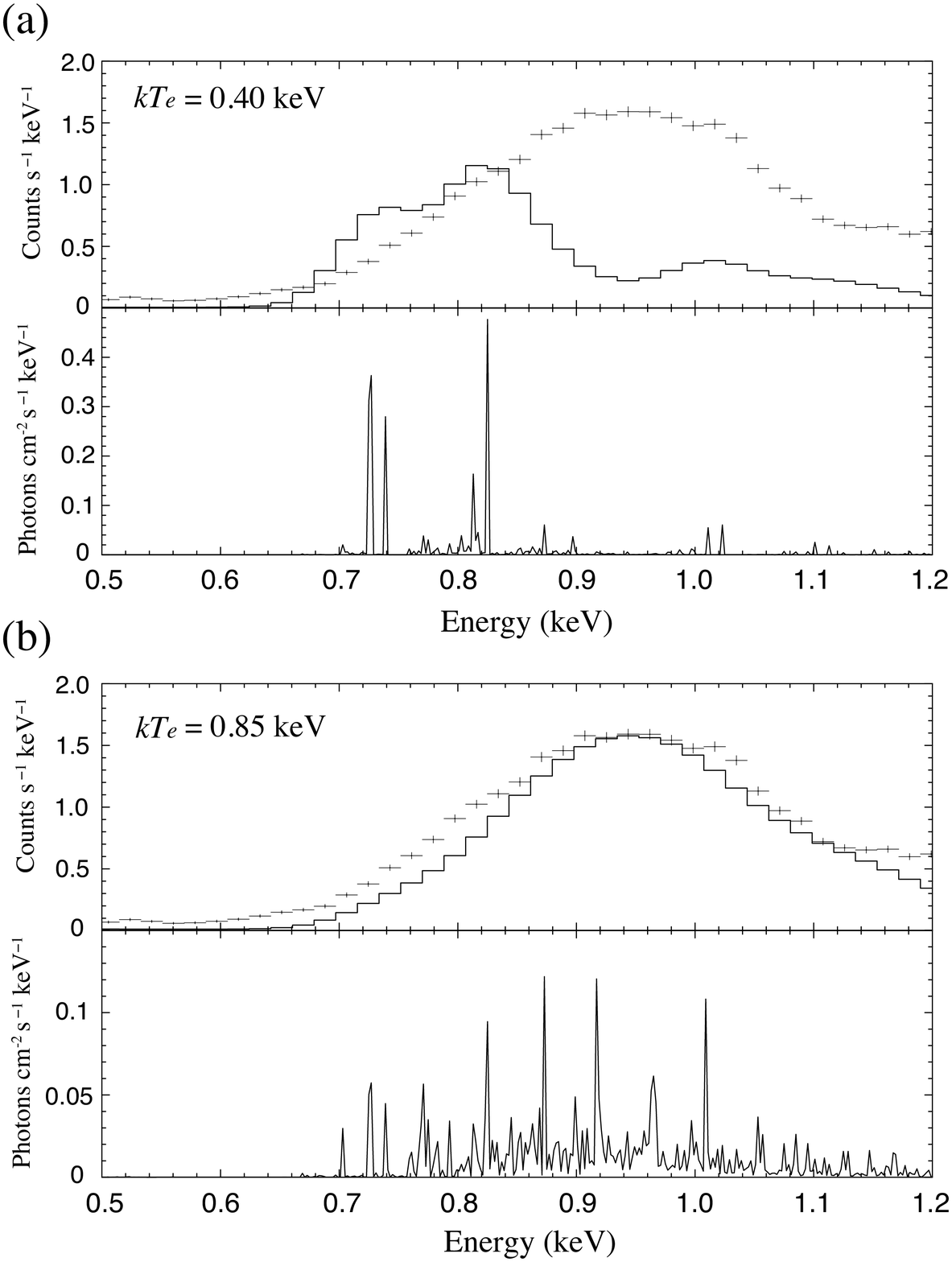} 
\end{center}
\caption{(a) NE spectrum of XIS0+3 (crosses) and the plasma model with (top) and without (bottom) response matrixes multiplied (solid lines) in the energy band of 0.5--1.2 keV.
The plasma model consists only of Fe and Ni components with $N_{\rm H}=1.7 \times 10^{21}~{\rm cm^{-2}}$, $kT_{e}=0.40~{\rm keV}$, $kT_{init}=3.0~{\rm keV}$ and $n_{e}t=6.0\times10^{11} {\rm cm^{-3}~s}$.
(b) Same as (a) but with $kT_{e}=0.85~{\rm keV}$.
}
\label{fig:fe_comp}
\end{figure}

We extrapolated the above RP model down to 0.5~keV.
Figure~\ref{fig:spectra}~(ne-iv) shows the model and the full-band spectrum in the energy band of 0.5--10.0~keV.
The spectral structure in the 0.7--1.1~keV band cannot be reproduced. 
With the electron temperature $kT_{e}=0.37~\rm keV$,
the model predicts a peak energy of the Fe and Ni L line complexes  at $\sim 0.8$~keV, which is significantly lower than the observed value of $\sim 0.95$~keV. 
Figure~\ref{fig:fe_comp} shows the NE spectrum of XIS0+3 and the plasma model in the energy band of 0.5--1.2 keV.
The model consists only of the Fe and Ni components with $N_{\rm H}=1.7 \times 10^{21}~{\rm cm^{-2}}$, $kT_{init}=3.0~{\rm keV}$ and $n_{e}t=6.0\times10^{11}~{\rm cm^{-3}~s}$.
The electron temperatures are 0.40~keV and 0.85~keV in Figures~\ref{fig:fe_comp} (a) and (b), respectively.
The dominant emission lines of the Fe and Ni components are Fe\emissiontype{XVII} 2p5 3s1$\rightarrow$2p6 (0.725~keV and 0.739~keV) and Fe\emissiontype{XVII} 2p5 3d1$\rightarrow$2p6 (0.812~keV and 0.826~keV) with $kT_e = 0.40$~keV, and Fe\emissiontype{XIX} 2p3 3s1$\rightarrow$2p4 (0.822~keV), Fe\emissiontype{XVIII} 2p4 3d1$\rightarrow$2p5 (0.873~keV), Fe\emissiontype{XX} 2p2 3d1$\rightarrow$2p3 (0.965~keV) and Fe\emissiontype{XXI} 2p1 3d1$\rightarrow$2p2 (1.009~keV) with $kT_e = 0.85$~keV.
The comparison between the data and the models indicates that the electron temperature of the Fe and Ni plasma should be $\sim 0.85$~keV in order to reproduce the observed structure.
On the other hand, the Si and S plasma demands lower $kT_{e}$ as we saw in the (ne-iii) fit.

The result indicates that the plasma has different $kT_{e}$ between Fe (+Ni) and the other elements.
\citet{Kamitsukasa2016} decomposed the ejecta of SNR G272.2$-$3.2 into components with different elements, considering that each element has different radial distributions.
Following their idea, we adopted a two-component VVRNEI model: one for Fe and Ni, and the other for lighter elements. 
The spectra can be well fitted with the model comprising a CIE for Fe and Ni and an RP for the other elements.
Figure~\ref{fig:spectra}~(ne-v) shows the spectrum with the best-fit model.
Although we tried IP and RP for the Fe and Ni component, the ionization parameter $n_{e}t$ cannot be constrained well.
Therefore, we here assumed the CIE model for simplicity. 

\section{Discussion}
\subsection{Spatial structure}
\begin{figure}
\begin{center}
 \includegraphics[width=7cm]{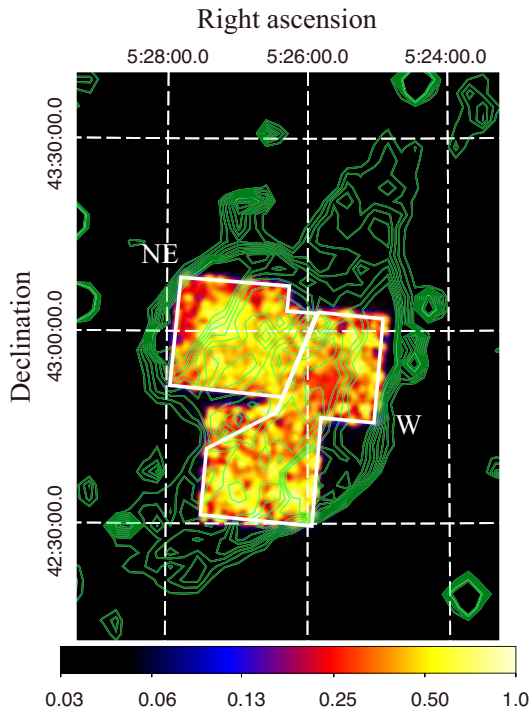} 
\end{center}
\caption{XIS images of G166.0+4.3 in 1.75--2.10 keV band (Top: Si-K band) and the 0.70--1.30 keV band (Bottom: Fe-L and Ni-L band) after subtraction of the NXB and correction of the vignetting effect. The images are divided by the corresponding underlying continuum image for which we selected the line-free band of 1.50--1.70~keV. 
The scale is normalized so that the peak becomes unity.
The contours are a radio image of the WENSS at 325 MHz.
}
\label{fig:img_band}
\end{figure}

The spectral analysis shows that the NE plasma contains two components; the low temperature RP component and the high temperature Fe-rich component.
This suggests that the distributions of the two components are different from each other.
In order to investigate the distribution, we made XIS images in the Si-K band (1.75--2.10~keV) and the Fe-L band (0.70--1.30~keV).
We divided them by the corresponding underlying continuum image for which we selected the line-free band of 1.50--1.70~keV.
We show the images in Figure~\ref{fig:img_band}, which indicates that Si distributes uniformly whereas the Fe distribution is more compact.
In some SNRs, regardless of the type of the explosion, distributions of Fe are observed to be centered (e.g., \cite{Uchida2009,Hayato2010}). 
The observed distributions are consistent with the widely believed picture of supernova nucleosynthesis that heavier elements tend to be produced closer to the center of the progenitor.

\begin{figure}
\begin{center}
 \includegraphics[width=7cm]{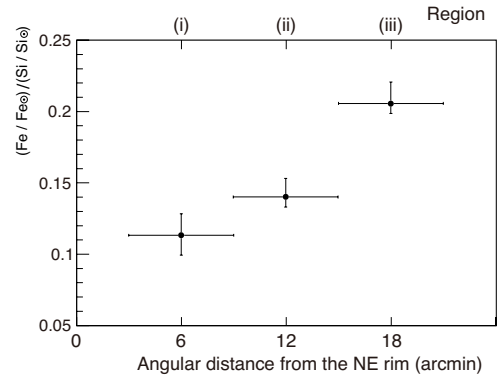} 
\end{center}
\caption{Abundance ratios of Fe to Si in the regions (i)--(iii) as a function of angular distances from the NE rim toward the center of the remnant.
}
\label{fig:fit_newregion}
\end{figure}

For a detailed analysis of the Fe distribution, we extracted spectra from the three regions (i)--(iii) in NE indicated by the white lines in the bottom panel of Figure~\ref{fig:img_band}.
We fitted the spectra of each region with the same model as Figure~\ref{fig:spectra}~(ne-v).
We fixed $kT_{e}$ of the RP component to the best-fit value, 0.46~keV.
Figure~\ref{fig:fit_newregion} shows the resultant abundance ratios of Fe to Si.
The ratio in the inner region is significantly higher and suggests the center-filled distribution of Fe.

\subsection{Ambient gas density and SNR age}
In our spectral analysis, the volume emission measures are $2.7 \times 10^{57}~\rm cm^{-3}$ and $1.0 \times 10^{58}~\rm cm^{-3}$ in W and NE, respectively.
The volumes are obtained to be $4.1 \times 10^{59}~f~\rm cm^3$ and $1.8 \times 10^{59}~f~\rm cm^3$ for the W and NE plasmas, respectively, where $f$ is the filling factor, assuming the morphologies to be prisms whose bases are the W and NE regions and depths are 40~pc and 20~pc for W and NE, respectively.
With the volume emission measures, we obtain the electron densities  $n_{e, \rm W} = 0.3~(f/0.1) ^{-1/2}~\rm cm^{-3}$ and $n_{e, \rm NE} = 0.9~(f/0.1) ^{-1/2}~\rm cm^{-3}$ for W and NE, respectively.
They suggest that the ambient gas is denser in NE than that in W.
This is supported by the radio morphology, the smaller shell in the eastern side. 
Another clue of denser gas in NE is the GeV gamma-ray emission detected in the eastern part of the remnant by Fermi LAT observations \citep{Araya2013}. 
If $\pi^0$ decays are the main production process for the gamma rays, dense gas, which acts as targets for accelerated protons for $\pi^0$ production, would be in NE.

In the spectral analysis, we obtained $n_{e, \rm W}~t = 4.0 \times 10^{11}~\rm cm^{-3}~s$ and $n_{e, \rm NE}~t = 6.1 \times 10^{11}~\rm cm^{-3}~s$ for the W and NE plasmas, respectively.
With $n_{e, \rm W}$ and $n_{e, \rm NE}$ calculated above based on the emission measures, the timescales are estimated to be $t = 1.4 \times 10^{5}~f^{1/2}~\rm yr$  for W and $t = 7.1 \times 10^{4}~f^{1/2}~\rm yr$ for NE.
The ionization timescale for the W plasma, which is an IP, is expected to be almost the same as the age of the remnant, and, therefore, we estimate the age to be $t_{age} = 4 \times 10^{4}~(f/0.1)^{1/2}~\rm yr$.
With $n_{e, \rm NE}~t$, we can estimate the elapsed time since recombination started to dominate over ionization to be $t_{rec} = 2 \times 10^{4}~(f/0.1)^{1/2}~\rm yr$ ($< t_{age}$).

\subsection{Origin of recombining plasma}
The rarefaction \citep{Itoh1989} and thermal conduction \citep{Kawasaki2002} scenarios have been considered for the formation process of RPs.
In either case, the environment of the remnant is a key to the RP production process.
We discovered an RP only in a part of G166.0+4.3, the outer part of the NE region, where the ambient gas is denser.
In the W region, the spectrum is reproduced by the IP model.
The different gas environments between the W and NE regions might be actually responsible for the difference of the plasma states.

In the rarefaction scenario, the electron temperature is decreased by adiabatic cooling when the blast wave breaks out of dense CSM into rarefied ISM.
The lower the ISM density is, the more effective the adiabatic cooling must be \citep{Shimizu2012}.
Therefore, in this scenario, we can naturally expect an RP in the W region whose ambient gas density would be lower.
Our result, where we found an RP only in the NE region, is not simply explained by the rarefaction senario. 

In the thermal conduction scenario, an RP is anticipated in a part of the remnant where blast waves are in contact with cool dense gas. 
In the case of G166.0+4.3, the RP was discovered indeed in the denser NE region.
In our results, the temperature of the inner Fe-rich plasma is higher ($kT_e = 0.87~\rm keV$) than that of the outer RP ($kT_e = 0.46~\rm keV$).
In this scenario, the plasma is cooled from the outside layers.
Therefore, it is possible that the plasma is cooled only in the outer part of the NE region.

\section{Summary}
We analyzed the Suzaku XIS data of SNR G166.0+4.3.
The results we obtained are summarized as follows.
\begin{enumerate}
\item The plasma in the NE region consists of two components. One is the Fe-rich component with $kT_e = 0.87_{-0.03}^{+0.02}~\rm keV$. 
The other is the RP component of the other lighter elements with $kT_e = 0.46 \pm 0.03~\rm keV$, $kT_{init} = 3~\rm keV (fixed)$ and $n_et = (6.1_{-0.4}^{+0.5}) \times 10^{11}~\rm cm^{-3}~s$.
\item The spectrum in the W region is well fitted with a one-component IP model with $kT_e = 0.83 \pm 0.01~\rm keV$, $kT_{init} = 0.01~\rm keV (fixed)$ and $n_et = (4.0_{-0.5}^{+0.7}) \times 10^{11}~\rm cm^{-3}~s$.
\item The Fe-rich plasma is concentrated near the center of the remnant whereas the other elements are distributed in the outer region. 
\item Our results of the spectral analysis are not simply explained by the rarefaction scenario, whereas the origin of the RP would be explained by the thermal conduction with the cool dense gas in the northeastern part of the remnant.

\end{enumerate}

\begin{ack}
The authors are grateful to Prof. Katsuji Koyama for helpful advice.
We deeply appreciate all the Suzaku team members.
We also thank the WENSS team.
This work is supported by JSPS Scientific Research grant numbers 15J01842 (H.M.), 15H02090, 26610047 and 25109004 (T.G.T.), 26800102 (H.U.), 25109004 (T.T.), and 16J00548 (K.K.N.).
\end{ack}

%
%



\end{document}